\documentclass[showpacs,twocolumn,prl,aps]{revtex4}
\usepackage{graphicx}

\begin{document}

\title[Short title for running header]{On the Origin of the Tunneling Asymmetry in the Cuprate Superconductors: a variational perspective}
\author{Hong-Yu Yang$^{1}$, Fan Yang$^{2}$, Yong-Jin Jiang$^{3}$ and  Tao Li$^{4}$}
\affiliation{$^{1}$Center for Advanced study, Tsinghua University
,Beijing 100084, P.R.China\\
$^{2}$Department of Physics, Beijing Institute of Technology
,Beijing 100081, P.R.China\\
$^{3}$Department of Physics, Zhejiang Normal University, Jinhua
321004, P.R.China\\
$^{4}$Department of Physics, Renmin University of China, Beijing
100872, P.R.China}
\date{\today}

\begin{abstract}
Through variational Monte Carlo calculation on Gutzwiller projected
wave functions, we study the quasiparticle(qp) weight for adding and
removing an electron from a high temperature superconductor. We find
the qp weight is particle-hole symmetric at sufficiently low energy.
We propose to use the tunneling asymmetry as a tool to study the
mechanism of electron incoherence in high temperature
superconductors.
\end{abstract}
\maketitle

The scanning tunneling microscopy(STM) plays an important role in
the study of the high temperature superconductors since it can
provide local information on the single particle properties with
ultrahigh energy resolution. A striking feature in the STM spectrum
of the high temperature superconductors is their remarkable
particle-hole asymmetry. The hole side of the spectrum always
dominate the particle side of the spectrum in hole doped
cuprates\cite{Pan}.

This asymmetry is not at all surprising if we take the high
temperature superconductors as a doped Mott insulators described by
the $t-J$ model. In such a doped Mott insulator, an added electron
has a reduced probability to contribute to the electron spectral
weight in the low energy subspace of no doubly occupancy. More
specifically, if the hole density in the system is $x$, then the
total spectral weight in the particle side of the spectrum is
reduced to $x,$ while the total spectral weight in the hole side of
the spectrum is not affected by the no double occupancy constraint.
Thus the total spectral weight is particle-hole asymmetric for small
$x$. However, such an asymmetry on the total spectral weight tell us
nothing about the distribution of the spectral weight at low energy.
To address the problem of tunneling asymmetry in the near vicinity
of the chemical potential, we need more detailed information on the
low energy excitation of the system.

Rantner and Wen addressed this problem in the slave-Boson mean field
theory of the $t-J$ model\cite{Rantner}. In their theory, the
tunneling asymmetry comes from the incoherent part of the electron
spectrum. In the slave-Boson mean field theory, an electron is split
into two parts, namely the Fermionic spinon part that carry spin and
the Bosonic holon part that carry charge. The superconducting state
is described by the Bose condensation of holon in the background of
BCS pairing of spinon. In the presence of the holon condensate, the
electron spectrum acquire a nonzero quasiparticle(qp) weight. In the
mean field theory, the spectral weight in the particle side of the
spectrum is totally coherent since the holon removed during the
particle injection process must originate from the holon condensate.
However, the hole side of the spectrum involves both coherent and
incoherent part since the holon injected during the particle
removing process can stay either in the condensate or out of it. In
the slave-Boson mean field theory, the qp weight differs from the
BCS result by a constant renormalization factor x, namely
$xu_{\mathrm{k}}^{2}$ for adding an electron and
$xv_{\mathrm{k}}^{2}$ for removing an electron. Thus if one neglect
the asymmetry due to the band structure effect near the chemical
potential, then the qp weight is particle-hole symmetric and
tunneling asymmetry must originate from the incoherent part of the
electron spectrum.

Recently, Anderson and Ong addressed the same problem with a
variational approach\cite{Anderson}. They constructed the
variational wave function for the ground state and the excited state
of the $t-J$ model following the original RVB idea. In their
treatment, the particle-hole asymmetry is taken into account
explicitly in the variational wave function by the introduction of a
fugacity factor $\mathrm{Z}(=\frac{2x}{1+x})$. This fugacity factor
is expected to play the role of the Gutzwiller projection into the
subspace of no double occupancy. In this theory, the electron
spectrum is dominated by the qp contribution and the tunneling
asymmetry originates from the coherent rather than the incoherent
part of the electron spectrum. Especially, when the excitation
energy considered is much larger than the pairing gap(so that the
particle-hole mixing due to the superconducting pairing is
negligible), the qp weight for adding an electron is a factor Z
smaller than that for removing an electron. In this theory, the qp
weight for adding an electron scales linearly with the hole density
$x$ near half filling , while the qp weight for removing an electron
scales linearly with $1-x$. The problem of tunneling asymmetry and
the qp weight is also addressed variationally in some other recent
works\cite{Randeria,Yunoki,Nave,Lee}. For example, it is proved by
Yunoki that the electron spectral weight in the particle side is
exhausted by the qp contribution for a superconductor described by
Gutzwiller projected BCS wave functions\cite{Yunoki}. However, a
clear understanding of the hole-like qp excitation of the Gutzwiller
projected state is still absent.

In this paper, we show through the variational Monte Carlo
calculation on Gutzwiller projected wave functions that the qp
weight in the t-J model is particle-hole symmetric at sufficiently
low energy. Especially, we find the qp weight for adding and
removing an electron in a projected Fermi sea state both converge to
the value determined from the jump of the momentum distribution
function on the Fermi surface. We find the qp weight for adding and
removing an electron both vanish at zero doping, as predicted by the
slave-Boson mean field theory. However, the qp weight calculated
from projected wave function show more complex momentum and doping
dependence. We find the qp weight vanishes like
$x^{\alpha_{\mathrm{k}}}$ near half filling. The exponent
$x^{\alpha_{\mathrm{k}}}$ increases monotonically from $\frac{1}{2}$
to 1 when one move from deep inside the Fermi surface to far outside
it in two spatial dimension. Thus the qp weight vanishes more slowly
near zero doping than that predicted by the slave-Boson mean field
theory. According to our result, the tunneling asymmetry near the
Fermi level should be attributed to the incoherent part of the
electron spectrum. We propose to use the tunneling asymmetry to
study the mechanism of electron incoherence in the high temperature
superconductors.

In the Landau theory of Fermi liquid, the quasiparticle plays a dual
role. On the one hand, the qp can be thought of as a particle-like
elementary excitation on the ground state of a $N$ particle system.
On the other hand, the qp can also be thought of as a constituent of
the grounds state of the $N+1$ particle system, provided that the qp
is located on the Fermi surface. The qp weight for adding an
electron into the system on the Fermi surface is thus equal to the
square of the matrix element of the electron creation operator
between the ground state of $N$ particle system and the ground state
of $N+1$ particle system,
\[
Z_{N}^{+}=\left\vert \left\langle g_{_{N+1}}\right\vert
c_{_{k}}^{\dagger }\left\vert g_{_{N}}\right\rangle \right\vert ^{2}
\]
while the qp weight for removing an electron from the system on the
Fermi surface is equal to the square of the matrix element of
electron annihilation operator between the ground state of $N$
particle system and the ground state of $N-1$ particle system,
\[
Z_{N}^{-}=\left\vert \left\langle g_{_{N-1}}\right\vert
c_{_{k}}\left\vert g_{_{N}}\right\rangle \right\vert
^{2}=Z_{N-1}^{+}
\]
In the thermodynamic limit, we have
\[
Z_{N}^{-}=Z_{N-1}^{+}\simeq Z_{N}^{+}
\]
thus the qp weight should be particle-hole symmetric. This simple
argument do not apply to the superconducting state. However, we do
not expect the superconducting pairing to change the conclusion. The
quasiparticle in the superconducting state is a mixture of particle
and hole components. Thus the superconducting pairing is expected to
enhance rather than reduce the particle-hole symmetry.

Now we calculate the qp weight variationally with the Gutzwiller
projected BCS-type wave functions. This kind of variational
description is widely used in the study of the t-J model and is
believed to capture the low energy physics of the
cuprates\cite{Zhang,Paramekanti,Shil,Sorella}, especially when the
doping is near optimal. The variational ground state, namely the
Gutzwiller projected BCS state with N particle is given by(in the
following we follow the notations of \cite{Yunoki})
\begin{equation}\label{1}
    |\Psi_{0}^{\mathrm{N}}\rangle=P_{\mathrm{N}}P_{\mathrm{G}}|\mathrm{BCS}\rangle,
\end{equation}
where $P_{\mathrm{N}}$ is projection onto the subspace of N
particles, $P_{\mathrm{G}}$ is the projection onto the subspace of
no double occupancy. $|\mathrm{BCS}\rangle$ denotes the unprojected
BCS mean field ground state. The elementary excitation above the
ground state can be similarly constructed by Gutzwiller projection
of BCS mean field excited state. For example, an particle-like
elementary excitation is constructed as follows
\begin{equation}\label{2}
    |\Psi_{\mathrm{k},\sigma}^{\mathrm{N}+1}\rangle=P_{\mathrm{N+1}}P_{\mathrm{G}}\gamma_{\mathrm{k},\sigma}^{\dagger}|\mathrm{BCS}\rangle
\end{equation}
where $\gamma_{\mathrm{k},\sigma}^{\dagger}$ is the creation
operator for Bogliubov quasiparticle on the BCS mean field ground
state which is related to the original electron operator
$c_{\mathrm{k},\sigma}^{\dagger}$ through the Bogliubov
transformation
\begin{equation}\label{3}
    \left(
    \begin{array}{c}
      \gamma_{\mathrm{k},\uparrow} \\
      \gamma_{-\mathrm{k},\downarrow}^{\dagger}
    \end{array}\right) =
    \left(
    \begin{array}{cc}
      u_{\mathrm{k}^{*}} & -v_{\mathrm{k}}^{*} \\
      v_{\mathrm{k}} & u_{\mathrm{k}}
    \end{array}\right)
    \left(
    \begin{array}{c}
      c_{\mathrm{k},\uparrow} \\
  c_{-\mathrm{k},\downarrow}^{\dagger}\end{array}\right)
\end{equation}
Similarly, a hole like elementary excitation with the same momentum
and spin is constructed as follows
\begin{equation}\label{4}
    |\Psi_{\mathrm{k},\sigma}^{\mathrm{N}-1}\rangle=P_{\mathrm{N-1}}P_{\mathrm{G}}\gamma_{\mathrm{k},\sigma}^{\dagger}|\mathrm{BCS}\rangle
\end{equation}
The qp weight for adding and removing an electron are given by the
overlap between the corresponding bare electronic states and the qp
states
\begin{equation}\label{5}
    Z_{\mathrm{k}}^{+}=
    \frac{|\langle\Psi_{\mathrm{k},\sigma}^{\mathrm{N+1}}|c_{\mathrm{k},\sigma}^{\dagger}|\Psi_{0}^{\mathrm{N}}\rangle|^{2}}
    {\langle\Psi_{\mathrm{k},\sigma}^{\mathrm{N+1}}|\Psi_{\mathrm{k},\sigma}^{\mathrm{N+1}}\rangle\langle\Psi_{0}^{\mathrm{N}}|\Psi_{0}^{\mathrm{N}}\rangle}
\end{equation}
and
\begin{equation}\label{6}
    Z_{\mathrm{k}}^{-}=
    \frac{|\langle\Psi_{\mathrm{k},\sigma}^{\mathrm{N-1}}|c_{\mathrm{-k},-\sigma}|\Psi_{0}^{\mathrm{N}}\rangle|^{2}}
    {\langle\Psi_{\mathrm{k},\sigma}^{\mathrm{N-1}}|\Psi_{\mathrm{k},\sigma}^{\mathrm{N-1}}\rangle
    \langle\Psi_{0}^{\mathrm{N}}|\Psi_{0}^{\mathrm{N}}\rangle}
\end{equation}

As pointed out by Yunoki, using the fact that
$P_{\mathrm{G}}c_{\mathrm{k},\sigma}^{\dagger}P_{\mathrm{G}}=P_{\mathrm{G}}c_{\mathrm{k},\sigma}^{\dagger}$\cite{Anderson}
, the qp weight for adding an electron can be related to the
momentum distribution function $n_{k}$ in the following way,
\begin{equation}\label{7}
    Z_{\mathrm{k}}^{+}=|u_{\mathrm{k}}|^{2}
    \frac{\langle\Psi_{\mathrm{k},\sigma}^{\mathrm{N+1}}|\Psi_{\mathrm{k},\sigma}^{\mathrm{N+1}}\rangle}
    {\langle\Psi_{0}^{\mathrm{N}}|\Psi_{0}^{\mathrm{N}}\rangle}
    = 1-\frac{\mathrm{N}_{\bar{\sigma}}}{L}-n_{\mathrm{k}}
\end{equation}
where $\frac{\mathrm{N}_{\bar{\sigma}}}{L}$ denotes the mean value
of the particle number with opposite spin. Thus to calculate
$Z_{\mathrm{k}}^{+}$, one only need to evaluate an ground state
expectation value. The calculation of $Z_{\mathrm{k}}^{+}$ is more
complex. However, using
$P_{\mathrm{G}}c_{\mathrm{k},\sigma}^{\dagger}P_{\mathrm{G}}=P_{\mathrm{G}}c_{\mathrm{k},\sigma}^{\dagger}$
and the Bogliubov transformation Eq.(4), we are able to show that
\begin{equation}\label{8}
    Z_{\mathrm{k}}^{-}=\alpha\frac{|u_{\mathrm{k}}(v_{\mathrm{k}}-u_{\mathrm{k}}\mathrm{O}_{\mathrm{k}})|^{2}}
    {Z_{\mathrm{k},N-2}^{+}}
\end{equation}
where the constant $\alpha$ is given by
\begin{equation}\label{9}
    \alpha=\frac{\langle\Psi_{0}^{\mathrm{N}}|\Psi_{0}^{\mathrm{N}}\rangle}
    {\langle\Psi_{0}^{\mathrm{N-2}}|\Psi_{0}^{\mathrm{N-2}}\rangle}
\end{equation}
and plays the role of the fugacity factor. $\mathrm{O}_{\mathrm{k}}$
is given by following overlap integral
\begin{equation}\label{10}
    \mathrm{O}_{\mathrm{k}}=\frac{\langle\Psi_{0}^{\mathrm{N}}|\Psi_{2\mathrm{k}}^{\mathrm{N}}\rangle}
    {\langle\Psi_{0}^{\mathrm{N}}|\Psi_{0}^{\mathrm{N}}\rangle}
\end{equation}
in which
\begin{equation}\label{11}
    |\Psi_{2\mathrm{k}}^{\mathrm{N}}\rangle=P_{\mathrm{N}}P_{\mathrm{G}}\gamma_{\mathrm{k},\uparrow}^{\dagger}
    \gamma_{\mathrm{-k},\downarrow}^{\dagger}|\mathrm{BCS}\rangle
\end{equation}

Further simplification is possible when there is no superconducting
pairing. In this case, one find
\begin{equation}\label{12}
    Z_{\mathrm{k}}^{+}=\frac{\langle\Psi_{\mathrm{k}}^{\mathrm{N+1}}|\Psi_{\mathrm{k}}^{\mathrm{N+1}}\rangle}
    {\langle\Psi_{0}^{\mathrm{N}}|\Psi_{0}^{\mathrm{N}}\rangle}
\end{equation}
for $\mathrm{k}$ outside the Fermi surface and
\begin{equation}\label{13}
    Z_{\mathrm{k}}^{-}=\frac{\langle\Psi_{0}^{\mathrm{N}}|\Psi_{0}^{\mathrm{N}}\rangle}
    {\langle\Psi_{\mathrm{k}}^{\mathrm{N-1}}|\Psi_{\mathrm{k}}^{\mathrm{N-1}}\rangle}
\end{equation}
for $\mathrm{k}$ inside the Fermi surface. Noticing the fact that
$|\Psi_{\mathrm{k}}^{\mathrm{N+1}}\rangle$
($|\Psi_{\mathrm{k}}^{\mathrm{N-1}}\rangle$) is nothing but the
variational ground state of the $N+1$($N-1$) system for $\mathrm{k}$
located on the Fermi surface , we have
\begin{equation}\label{14}
    Z_{\mathrm{k_{F}^{+}}}^{+}=\frac{\langle\Psi_{0}^{\mathrm{N+1}}|\Psi_{0}^{\mathrm{N+1}}\rangle}
    {\langle\Psi_{0}^{\mathrm{N}}|\Psi_{0}^{\mathrm{N}}\rangle}
\end{equation}
\begin{equation}\label{15}
    Z_{\mathrm{k_{F}^{-}}}^{-}=\frac{\langle\Psi_{0}^{\mathrm{N}}|\Psi_{0}^{\mathrm{N}}\rangle}
    {\langle\Psi_{0}^{\mathrm{N-1}}|\Psi_{0}^{\mathrm{N-1}}\rangle}
\end{equation}
where $\mathrm{k_{\mathrm{F}}^{\pm}}$ denotes momentum located on
the Fermi surface. Thus, if the qp weight is a continuous function
of particle number, it should be particle-hole symmetric in the
thermodynamic limit. Furthermore, using Yunoki's relation and the
fact that $Z_{\mathrm{k}}^{+}$ vanish for $\mathrm{k}$ inside the
Fermi surface, we have
\begin{equation}\label{16}
    n_{\mathrm{k}}=1-\frac{n}{2}
\end{equation}
for $\mathrm{k}$ inside the Fermi surface and
\begin{equation}\label{17}
    Z_{\mathrm{k_{F}^{+}}}^{+}=\Delta n_{\mathrm{k_{F}}}
\end{equation}
where $\Delta n_{\mathrm{k_{F}}}$ denotes the jump of
$n_{\mathrm{k}}$ on the Fermi surface. Thus both
$Z_{\mathrm{k}}^{+}$ and $Z_{\mathrm{k}}^{-}$ converge to $\Delta
n_{\mathrm{k_{F}}}$ on the Fermi surface in the thermodynamic limit,
consistent with a Fermi liquid interpretation of the quasiparticle
excitation.

We now present the result of VMC calculation. Fig.(1) shows the
momentum distribution function and the qp weight for the Gutzwiller
projected Fermi sea on a $18\times18$ lattice. Here the mean field
Fermi sea is generated by a nearest neighboring hopping term on a
square lattice with periodic-periodic boundary condition. The hole
number is kept at 42($x\simeq0.13$) so that the closed-shell
condition is satisfied. As shown in Fig.1, the total qp weight is a
continues function of momentum across the Fermi surface. Thus the qp
weight must be particle-hole symmetric on the Fermi surface in the
thermodynamic limit, as we have argued before(note that the qp
weight for adding(removing) an electron vanishes inside(outside) the
Fermi surface). Note also that at this doping level the qp weight
for removing an electron is only slightly higher than that for
adding an electron in the whole Brillouin zone. Thus the
quasiparticle contribution to the STM spectrum should be
particle-hole symmetric at energy scale much smaller the band width.

\begin{figure}[h!]
\includegraphics[width=9cm,angle=0]{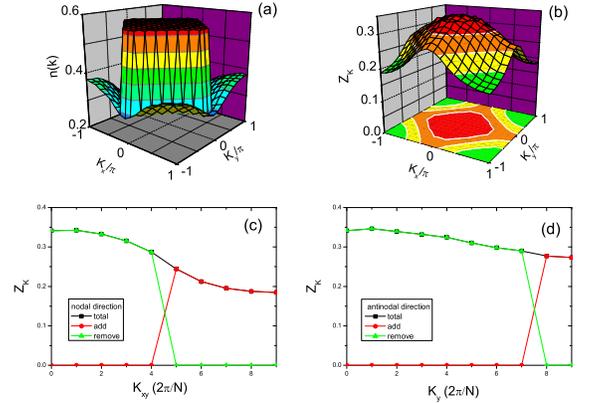}
\caption{VMC result for Gutzwiller projected Fermi sea on a
$18\times18$ lattice with 42 holes$(x\simeq0.13)$. The hole density
is choosen that the closed shell condition is satisfied for system
with periodic-periodic boundary condition. The mean field state is
generated by nearest-neighbouring hopping term on the lattice. (a)
momentum distribution function. (b)total qp weight.(c)qp weight in
the $(0,0)-(\pi,\pi)$ direction.(d)qp weight in the $(0,0)-(0,\pi)$
direction.} \label{fig1}
\end{figure}

Another implication of the above result is that most spectral weight
in the hole side of the spectrum is incoherent. To show this more
clearly, we plot in Fig.(2) the doping dependence of the
$Z_{\mathrm{k}=(0,0)}^{-}$ and $Z_{\mathrm{k}=(\pi,\pi)}^{+}$. From
the figure we see that both $Z_{\mathrm{k}=(0,0)}^{-}$ and
$Z_{\mathrm{k}=(\pi,\pi)}^{+}$ vanish near half filling, in
agreement with the slave-Boson mean field theory prediction. As
compared to the slave-Boson mean field result( $Z_{\mathrm{k}}^{+}=x
u_{\mathrm{k}}^{2}$,$ Z_{\mathrm{k}}^{-}=xv_{\mathrm{k}}^{2}$), the
qp weight calculated from projected wave function show more complex
momentum dependence. The origin of the momentum dependence can be
traced back to the non-monotonic behavior of $n(\mathrm{k})$. As
first noted by in \cite{Yokoyama}, the non-monotonic behavior of
$n(\mathrm{k})$ comes from the correlated hoping nature of the
electron in the t-J model. Another difference between the
slave-Boson mean field theory and the projected wave function is the
doping dependence of the qp weight(although they both vanish at half
filling). In the slave-Boson mean field theory, the qp weight scales
linearly with $x$ at all momentums. However, we find the qp weight
calculated from the projected wave function show
$x^{\alpha_{\mathrm{k}}}$ behavior at low doping, where
$\alpha_{\mathrm{k}}$ is a momentum dependent exponent. For the two
dimensional projected Fermi sea, $\alpha_{\mathrm{k}}$ increase
monotonically from $\frac{1}{2}$ deep inside the Fermi surface to 1
far outside the Fermi surface. Thus the qp is more robust in the
variational description than in the slave-Boson mean field theory.
This difference may be caused by spin-charge recombination induced
by the Gutzwiller projection\cite{Yang}.

\begin{figure}[h!]
\includegraphics[width=9cm,angle=0]{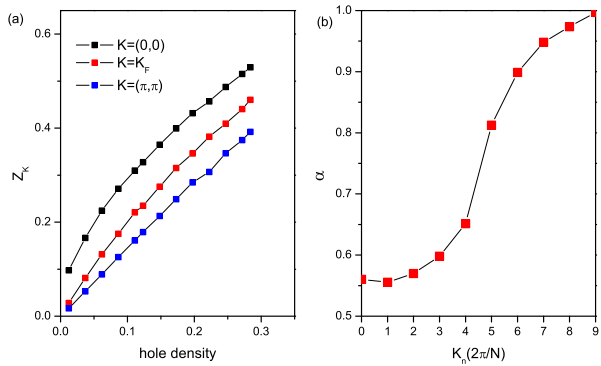}
\caption{Doping dependence of the qp weight as calculated from the
Gutzwiller projected Fermi sea on a $18\times18$ lattice as a
function of hole density. Periodic-antiperiodic boundary condition
is used in the calculation to maximize the number of hole density
that satisfy the closed-shell condition. (a)doping dependence of
$Z_{\mathrm{k}}$ at three momentums along $(0,0)$ to $(\pi,\pi)$,
namely $\mathrm{k}=(0,0),\mathrm{k}_{F} and (\pi,\pi)$(note that
since the boundary condition is periodic-antiperiodic, the momentums
are not exactly in the nodal direction).(b)momentum dependence of
the exponent $\alpha_{\mathrm{k}}$ in the nodal direction.}
\label{fig1}
\end{figure}

Next we present the result for the Gutzwiller projected d-wave BCS
state. The mean field state is generated by the following
Hamiltonian
\begin{eqnarray*}
  \mathrm{H}_{MF} &=& -\sum_{<ij>,\sigma}(c_{i\sigma}^{\dagger}c_{j\sigma}+h.c.)-\mu\sum_{i\sigma}c_{i\sigma}^{\dagger}c_{i\sigma} \\
   &+& \Delta\sum_{<ij>}d_{ij}(c_{i\uparrow}^{\dagger}c_{j\downarrow}^{\dagger}+c_{j\uparrow}^{\dagger}c_{i\downarrow}^{\dagger}+h.c.)
\end{eqnarray*}
in which $d_{ij}$ is the form factor for d-wave pairing and the sum
is limited to nearest-neighboring sites. We take $\Delta=0.1$ and
the hole number is kept at 42. $\mu$ is determined by the mean field
equation for density. Fig.(3) shows the momentum distribution
function and the qp weight on a $18\times18$ lattice with
periodic-periodic boundary condition. The result is basically the
same as that of the projected Fermi sea state apart from the
particle-hole mixture in the vicinity of the Fermi surface.  One
thing to note is that the total qp weight is a monotonic function of
momentum in the nodal direction and do not exhibit pocket structure
of the kind found for the $Z_{\mathrm{k}}^{+}$ in \cite{Nave}. To
clarify the situation, we plot the momentum dependence of
$Z_{\mathrm{k}}^{+}$ and $Z_{\mathrm{k}}^{-}$ separately in Fig.(4).
While a well defined pocket structure is observed in
$Z_{\mathrm{k}}^{+}$, no corresponding structure exist in
$Z_{\mathrm{k}}^{+}$. The origin of the pocket structure in
$Z_{\mathrm{k}}^{+}$ can be traced back to the non-monotonic
behavior of $n(\mathrm{k})$ around the nodal point as a result of
the correlated hopping of electron in the t-J model and the d-wave
pairing. However, $Z_{\mathrm{k}}^{-}$ do not show such pocket
structure since it is a monotonically decreasing function of
momentum in the nodal direction. At the same time, since the
decrease of $Z_{\mathrm{k}}^{+}$ in the $(0,\pi)$ to $(\pi,0)$
direction due to pairing is compensated by the increase of
$Z_{\mathrm{k}}^{-}$, the total qp weight do not show pocket
structure.
\begin{figure}[h!]
\includegraphics[width=9cm,angle=0]{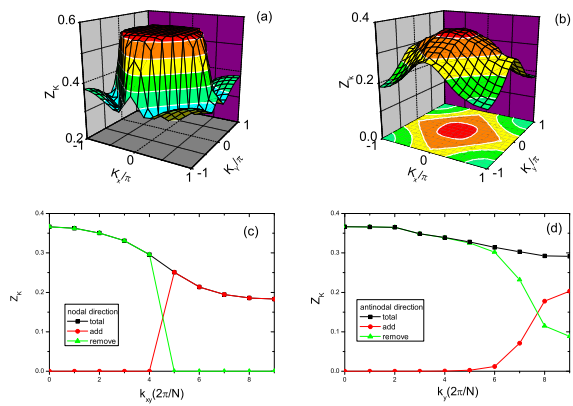}
\caption{VMC result for Gutzwiller projected d-wave BCS state on a
$18\times18$ lattice with 42 holes$(x\simeq0.13)$ and
$\frac{\Delta}{t}=0.1$. (a) momentum distribution function. (b)total
qp weight.(c)qp weight in the $(0,0)-(\pi,\pi)$ direction.(d)qp
weight in the $(0,0)-(0,\pi)$ direction.} \label{fig3}
\end{figure}

\begin{figure}[h!]
\includegraphics[width=9cm,angle=0]{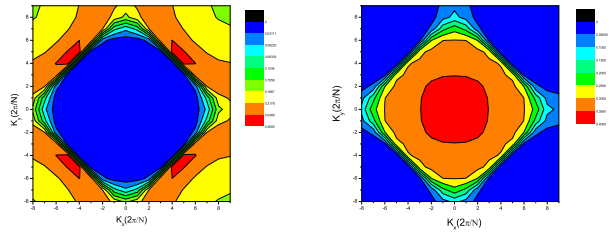}
\caption{Momentum dependence of $Z_{\mathrm{k}}^{+}$ and
$Z_{\mathrm{k}}^{-}$ in the Gutzwiller projected d-wave BCS state.}
\label{fig4}
\end{figure}

Above we have show that qp weight is particle-hole symmetric at
sufficiently low energy. Thus the tunneling asymmetry at low energy
should be attributed to the incoherent part of the electron
spectrum. Since the electronic spectrum in the particle side is
totally coherent, by subtracting the STM spectrum on the hole side
from that on the particle side we can extract the incoherent part of
the electron spectrum. For strongly correlated system like cuprates,
the information on the incoherent spectral weight is of great value
. Through studying this spectrum, one can figure out the mechanism
by which the bare electron decay into the many particle excitations
and also the nature of the many particle excitations itself. In the
context of the cuprates, two hotly discussed mechanisms to generate
electron incoherence are scattering with some bosonic collective
mode(like the neutron resonance mode)\cite{Eschrig} and electron
fractionalization as we have discussed in the slave-Boson mean field
theory\cite{Rantner}. It is interesting to see which mechanism
dominate the tunneling asymmetry at low energy.

The author would like to thank members of the HTS group at CASTU for
discussion. H.Y. Yang and T. Li is supported by NSFC Grant No.
90303009.

\end{document}